\documentclass[12pt,preprint]{aastex}

  \def\kms{km s$^{-1}$} 
  
  \def\and{$\&$ }

\begin{document} 
\setcounter{equation}{0}

  \title{The Environment of Local Ultraluminous Infrared Galaxies}
  \author{B. A. Zauderer\altaffilmark{1}, S. Veilleux\altaffilmark{1}, and H.K.C. Yee\altaffilmark{2}}
  \altaffiltext{1}{Department of Astronomy, University of Maryland,
    College Park, MD 20470; zauderer@mail.umd.edu, veilleux@astro.umd.edu}
  \altaffiltext{2}{Department of Astronomy and Astrophysics,
    University of Toronto, Toronto, ON M5S 3H4, Canada; hyee@astro.utoronto.ca}

\begin{abstract}   

  The spatial cluster-galaxy correlation amplitude, $B_{gc}$, is
  computed for a set of 76 $z < 0.3$ ultraluminous infrared galaxies
  (ULIRGs) from the 1-Jy sample. The $B_{gc}$ parameter is used to
  quantify the richness of the environment within 0.5 Mpc of each
  ULIRG. We find that the environment of local ULIRGs is similar to
  that of the field with the possible exceptions of a few objects with
  environmental densities typical of clusters with Abell richness
  classes 0 and 1. No obvious trends are seen with redshift, optical
  spectral type, infrared luminosity, or infrared color
  ($f_{25}/f_{60}$). We compare these results with those of local AGNs
  and QSOs at various redshifts. The 1-Jy ULIRGs show a broader range
  of environments than local Seyferts, which are exclusively found in
  the field. The distribution of ULIRG $B_{gc}$-values overlaps
  considerably with that of local QSOs, consistent with the scenario
  where some QSOs go through a ultraluminous infrared phase. However,
  a rigorous statistical analysis of the data indicates that these two
  samples are not drawn from the same parent population.  The $B_{gc}$
  distribution of QSOs shows a distinct tail at high $B_{gc}$-values
  which is not apparent among the ULIRGs. This difference is
  consistent with the fact that some of the QSOs used for this
  comparison have bigger and more luminous hosts than the 1-Jy ULIRGs.

\end{abstract}


\section{Introduction} 

Ultraluminous Infrared Galaxies (ULIRGs) are defined as galaxies with
L$_{\rm IR}$ = L(8 $-$ 1000 $\mu$m) $\ge$ 10$^{12}$L$_{\odot}$ (see
reviews by Sanders \& Mirabel 1996; Lonsdale, Farrah, \& Smith 2006).
This luminosity limit is roughly equivalent to the minimum bolometric
luminosity of QSOs. At luminosities above 10$^{12}$L$_{\odot}$, the
space density of ULIRGs in the local universe is greater than that of
optically selected quasars with similar bolometric luminosities by a
factor of $\sim$ 1.5. Thus local ULIRGs represent the most common type
of ultraluminous galaxy.  Systematic optical and near-infrared imaging
surveys have revealed that local ULIRGs are almost always undergoing
major mergers (e.g., Surace \& Sanders 1999; Surace, Sanders, \& Evans
2001; Scoville et al.\ 2000; Veilleux et al.\ 2002, 2006). Most of the
gas and star formation (and AGN) activity in these systems are
concentrated well within the central kpc (e.g., Downes \& Solomon
1998; Soifer et al. 2000, 2001).  Ground-based optical and
near-infrared spectroscopic studies of these objects have shown that
at least 25\% -- 30\% of them show genuine signs of AGN activity
(e.g., Kim, Veilleux, \& Sanders 1998; Veilleux, Kim, \& Sanders 1997,
1999; Veilleux, Sanders, \& Kim 1999).  This fraction increases to
$\sim$ 50\% among the objects with log[$L_{\rm IR}/L_\odot$] $\ga$
12.3. These results are compatible with those from mid-infrared
spectroscopic surveys (e.g., Genzel et al. 1998; Lutz et al. 1998;
Lutz, Veilleux, \& Genzel 1999; Rigopoulou et al.  1999; Tran et al
2001).

ULIRGs are relevant to a wide range of astronomical issues, including
the role played by galactic mergers in forming some or all elliptical
galaxies (Genzel et al.\ 2001; Veilleux et al.\ 2002), the efficiency
of transport of gas into the central regions of such mergers and the
subsequent triggering of circumnuclear star formation (e.g., Mihos \&
Hernquist 1996; Barnes 2004), the resulting heating and metal
enrichment of the IGM by galactic winds (e.g.  Rupke, Veilleux, \&
Sanders 2002, 2005ab; Veilleux, Cecil, \& Bland-Hawthorn 2005; Martin
2005), the potential growth and fueling of supermassive black holes
and the possible origin of quasars (Sanders et al. 1988). The
discovery of $z = 1 - 4$ submm sources with SCUBA (e.g., Smail et al.\
1997; Hughes et al.\ 1998) suggests that ULIRGs are also relevant to
the dominant source of radiant energy in the universe today.  Indeed,
integration of the light from the SCUBA population shows that it may
account for most of the submm/far-infrared background, as a
result of the strong cosmological evolution of these sources (e.g.,
Chapman et al.\ 2005). Thus, while the present-day ULIRGs provide a
relatively small contribution to the total present background, their
cousins at high $z$ are fundamentally important in this regard.

If ULIRGs are the predecessors of QSOs, one would expect ULIRGs and
QSOs to live in similar environments. Surprisingly little has been
published on the environments of local ULIRGs, in stark contrast to
the abundant literature on the small- and large-scale environments of
AGNs and QSOs (e.g., Yee, Green, \& Stockman 1986; Yee \& Green 1987;
Ellingson et al.\ 1991; Hill \& Lilly 1991; de Robertis et al.\ 1998;
McLure \& Dunlop 2001; Wold et al.\ 2000, 2001; Barr et al.\ 2003;
Miller et al.\ 2003; Kauffmann et al.\ 2004; S\"ochting et al.\ 2004;
Wake et al.\ 2004; Croom et al.\ 2005; Waskett et al.\ 2005; Serber,
Bahcall, \& Richards 2006) and the growing literature on the
environments of $z \ga 1$ ULIRGs (e.g., Blain et al.\ 2004; Farrah et
al.\ 2004, 2006).  To our knowledge, Tacconi et al.\ (2002) is the
only published study that has attempted to quantify the environments
of local ULIRGs. They correlated the positions of local ULIRGs with
the catalogs of galaxy clusters and groups available in NED and found
that none of them are located within a galaxy cluster.  The lack of
comprehensive imaging database at the time prevented them from
carrying out a more quantitative clustering analysis of these objects.

The present paper remedies the situation by using the large imaging
database of Veilleux et al.\ (2002) to quantify the environment of
local ($\langle z \rangle \sim 0.15$) ULIRGs from the 1-Jy sample.  We
note that the spectroscopy portion of the Sloan Digital Sky Survey
(SDSS) provides redshift information for only the bright tail of the
galaxy luminosity function at $z \sim 0.15$, so a method that relies
solely on the photometric measurements of the galaxies in the field
surrounding the ULIRG must be used for the present analysis.  The
properties of the 1-Jy sample and imaging dataset are reviewed in \S
2.  In \S 3, the procedure for deriving the environmental richness,
$B_{gc}$, is outlined.  Results for our sample are presented in \S 4.
The findings of environmental studies for quasars and Seyferts are
compared with our results in \S 5. Our conclusions are summarized in
\S 6. We use $H_0$ = 50 km s$^{-1}$ Mpc$^{-1}$, $\Omega_m = 1$, and
$\Omega_{\lambda}$ = 0 throughout this paper.  These values were
selected to match those of previous studies and facilitate
comparisons; they have no effect on our conclusions.

\section{Sample}

The {\em IRAS} 1-Jy sample of 118 ULIRGs identified by Kim \& Sanders
(1998) is the starting point of our investigation.  The 1-Jy ULIRGs
were selected to have high galactic latitude ($\vert b \vert$ $\ge$
30$^\circ$), 60-$\mu$m flux greater than 1 Jy, 60-$\mu$m flux greater than
their 12-$\mu$m flux (to exclude infrared-bright stars), and ratios of
60-$\mu$m flux to 100-$\mu$m flux above $10^{-0.3}$ (to favor the
detection of high-luminosity objects).
 
All 1-Jy ULIRGs were imaged at optical (R) and near-infrared
(K$^\prime$) wavelengths using the U.\ of Hawaii 2.2-meter telescope.
The present study uses only the R-band images since they have a larger
field of view (FOV) and are deeper than the K$^\prime$-band images.
The R filter at 6400 \AA\ was a Kron-Cousins filter.  Details of the
observations and data reduction can be found in Kim, Veilleux, \&
Sanders (2002). The analysis of these data is presented in Veilleux et
al.\ (2002). These data are part of comprehensive imaging and
spectroscopic surveys which also include a large set of optical and
near-infrared spectra of the nuclear sources (Veilleux et al. 1999ab
and references therein), a growing set of spatially-resolved
near-infrared spectra to study the gas and stellar kinematics of the
hosts (Genzel et al.\ 2001; Tacconi et al.\ 2002; Dasyra et al.\
2006a, 2006b), and mid-infrared spectra from the Infrared Space
Observatory (ISO) and the {\em Spitzer} Space Telescope (SST; e.g.,
Genzel et al. 1998; Veilleux et al. 2006b, in prep.). This effort is
called {\em QUEST}: Quasar / ULIRG Evolutionary Study.

Since the set of data presented in Kim et al.\ (2002) was compiled
from observations made over the course of 14 years, a variety of CCDs
were used and the FOV sizes and spatial resolutions are not uniform.
For consistency, we limit the set of data in this paper to the images
of the 76 objects taken under good photometric conditions with the TEK
2048 $\times$ 2048 CCD.  Of these images, 32 (42\%) were irrecoverably
cropped during an earlier stage of data reduction and have a
significantly reduced FOV size.  The effects of the cropping on the
results of our analysis are discussed in \S 4.  Table 1 lists the
objects in our sample along with the FOV size and several other
properties of the sources.  

\section{Analysis}

In this section we explain the methods that we used to quantify the
environment richness around each ULIRG. First, we describe the
algorithms used to find objects in the field and identify them as
stars or galaxies.  Next, we discuss the formalism applied to
calculate the environment richness parameter, $B_{gc}$. The techniques
used for our analysis have already been described in detail in Yee
(1991), Ellingson et al.\ (1991), Yee \& Lopez-Cruz (1999), and
Gladders \& Yee (2005); here we highlight the main steps.

\subsection{Object Identification and Classification}

Object identification was accomplished using the Picture Processing
Package (PPP) developed by one of us (Yee 1991).  This program
systematically examines each pixel in the image and determines whether
it has the potential to be part of an object: a star, a galaxy, a
cosmic ray, or an artifact of the CCD.  After running through a series
of tests, the PPP object finding program identifies and catalogs the
location and peak brightness of objects in the image.  The algorithms
used here are modified versions of that used by Kron (1980), which
depend on searching for local maxima. They have been shown to be
robust for object identification in sparse to moderately crowded
fields (Yee 1991).  The 1-Jy sample selection criterion $\vert b \vert
\ge 30^\circ$ avoids extremely crowded fields (and reduces the effects
of dust extinction on the galaxy counts), which could lead to object
misclassification and erroneous environment richness measurements.  We
therefore find that this object finding routine is perfectly adequate
for all ULIRGs in our sample

To address the problem of bad pixels or cosmic rays, objects were
thrown out automatically if a given pixel was five times brighter than
those immediately surrounding it.  This did not always work well
because bright bad pixels are sometimes surrounded by other bad pixels.
So, some misidentified objects were also identified by eye and removed
by hand.

The next step was to run an aperture photometry algorithm on the
identified objects in each image to determine whether these objects
are stars or galaxies.  For each object, a growth curve was calculated
using a series of circular apertures centered on the intensity
centroid of the object.  A reference-star growth curve was created for
each quadrant of the CCD frame by averaging the growth curves of
bright, isolated, and unsaturated stars within each quadrant. The
growth curves of the other objects were then compared with the
reference-star growth curve using the classification parameter $C_2$
defined by Yee (1991). In essence, $C_2$ computes the average
difference per aperture between the growth curves of the objects and
the growth curve of the reference star after they have been scaled to
match at the center and effectively compares the ratio between the
fluxes in the center and the outer part of an object with that of the
reference star. This method has been thoroughly tested by Yee (1991);
readers interested in knowing more about this classification scheme
should refer to this paper for detail.

\subsection{Environment Richness Parameter}

We use the parameter $B_{gc}$ to quantify the richness of the
environment of ULIRGs. $B_{gc}$ is the amplitude of the galaxy-galaxy
correlation function calculated for each object of interest
individually.  It was first used by Longair \& Seldner (1979) to
measure the environment of radio galaxies using galaxy counts, and
subsequently adopted in most studies of the environments of quasars
and other active galaxies (e.g., Yee \& Green 1984; Ellingson et al.\
1991; de Robertis et al.\ 1998; McLure \& Dunlop 2001; Wold et al.\
2000, 2001; Barr et al.\ 2003; Waskett et al.\ 2005), and also used as
a quantitative measurement of galaxy cluster richness (e.g., Andersen
\& Owen, 1994; Yee \& L\'opez-Cruz 1999).  Yee \& Lopez-Cruz (1999)
have demonstrated the robustness of the $B_{gc}$ parameter when
galaxies are counted to different radii and to different depth.
Furthermore, measurements of the environmental richness based on the
photometrically-derived $B_{gc}$-values have been shown to be entirely
consistent with measurements based on spectroscopic data.  This was
demonstrated by Yee \& Ellingson (2003), who used the data from the
Canadian Network for Observational Cosmology Cluster Redshift Survey
(CNOC1) to compare $B_{gc}$-values derived from (1) photometric data
with background subtraction, and (2) from properly weighted
spectroscopy data to account for incompleteness. We describe briefly
the procedure for deriving $B_{gc}$ below.

In order to determine the richness of the environment around a ULIRG,
we need to count the number of galaxies within a spherical volume with
radius, $r$, from the ULIRG of interest.  However, we necessarily must
begin with a two-dimensional image, which is a projection of this
volume onto the sky plane. The number of galaxies in a solid angle
$d\Omega$, at an angular distance $\theta$ from the object of
interest is given by (Seldner \& Peebles 1978)
\begin{equation}
N(\theta)d\Omega = N_g[1 + \omega(\theta)]d\Omega,
\end{equation}
where $N_g$ is the average surface density of galaxies and
$w(\theta)$, the angular correlation function, can be expressed
approximately as a power law, 
\begin{equation}
w(\theta) = A_{gc}\theta^{1-\gamma}.
\end{equation}
$A_{gc}$ is a measure of the average enhancement of galaxies in
angular area, and $\gamma \approx 1.77$ empirically.  Integrating
equation (2) within a circle with radius $\theta$ yields
\begin{equation}
A_{gc} = \frac{N_{tot} - N_{bgc}}{N_{bgc}}\frac{(3-\gamma)}{2}\theta^{\gamma-1},
\end{equation}
where $N_{tot}$ and $N_{bgc}$ are the the total numbers of galaxies and
background galaxies, respectively, within an angular
radius of $\theta$.

Next, the two-dimensional parameters must be translated into three
dimensions.  The angular correlation function $w(\theta)$ is translated
into the spatial correlation function, $\xi(r)$, which describes the
number of galaxies in volume element $dV$ at distance $r$ from the
object of interest.  It can be shown that $\xi(r) = B_{gc}
r^{-\gamma}$, where $\gamma$ has the same value as in equation (3) and
$B_{gc}$ is the spatial correlation amplitude, a measure of the
richness of the environment around the galaxy. Longair \& Seldner
(1979) have shown that
\begin{equation}
B_{gc} = \frac{A_{gc} n_{bg}(m) D^{\gamma-3}}{I_{\gamma}\Psi(m,z)}.
\end{equation}
The constant I$_{\gamma}$ is an integration constant which depends on
$\gamma$ (Groth \& Peebles 1977).  $n_{bg}(m)$ is the expected count
per unit angular area
of background galaxies brighter than apparent magnitude $m$,
$\Psi(m,z)$ is the normalized integrated luminosity function of
galaxies to apparent magnitude $m$, at redshift $z$ of the cluster,
and $D$ is the angular diameter distance to the ULIRG at redshift $z$.
For our calculations, we used $\gamma$ = 1.77, $I_{\gamma}$ = 3.78,
and the cosmological parameters H$_{0}$ = 50 \kms Mpc$^{-1}$,
$\Omega_m$ = 1, and $\Omega_{\lambda}$ = 0 to match those of previous
papers.  For $\Psi$ and $n_{bg}(m)$, we use the luminosity function
and background counts derived from the Red-Sequence Cluster Survey
(RCS; e.g., Gladders \& Yee 2005).

The uncertainty on B$_{gc}$ is computed using the formula
\begin{equation}
\frac{\Delta B_{gc}}{B_{gc}} = \frac{(N_{net} + 1.3^2 N_{bg})^{\onehalf}}{N_{net}}
\end{equation}
where $N_{net}$ is the net counts of galaxies over the background
counts, $N_{bg}$. This is a conservatively large error estimate as it
includes the expected counting statistics in $N_{net}$ and the
expected dispersion in background counts. The factor 1.3$^2$ is
included to account approximately for the additional fluctuation from
the clustered (and hence non-Poissonian) nature of the background
counts (discussed in detail in Yee, Green, \& Stockman 1986).  We
follow the prescription of Yee \& L\'opez-Cruz (1999) and integrate
the luminosity function from approximately $M_R = -25$ to $M^*_R + 2$
(where $M^*_R \approx -22.3$ for our cosmology) to calculate the
galaxy counts. This corresponds roughly to R = 15 -- 20 for the
galaxies in our sample ($\langle z \rangle \approx 0.15$).  This range
of integration was found by Yee \& L\'opez-Cruz (1999) to reduce the
sensitivity to small intrinsic variation of $M^*$ and variations in
the faint-end slopes of the cluster luminosity function.  The $B_{gc}$
parameter is computed over a radius $r = 500$ kpc, either directly
from the data when FOV $\ge$ 1 Mpc or extrapolated to this radius when
FOV $<$ 1 Mpc. This radius is selected to match that of previous
  studies.  The $B_{gc}$ parameter is not sensitive to this radius
  (Yee \& Lopez-Cruz 1999; see also \S 4).

\section{Results}

The spatial correlation amplitude parameter, B$_{gc}$, was computed for
each of the 76 ULIRGs in our sample.  The B$_{gc}$-values and
associated 1-$\sigma$ uncertainties are listed for each object in
Table 1.  The average (median) value of $B_{gc}$ and 1-$\sigma$
scatter around the mean for our sample of 76 ULIRGs is $\langle B_{gc}
\rangle = 35 \pm 198$ Mpc$^{1.77}$ ($-3$ Mpc$^{1.77}$).  For
comparison, the $B_{gc}$-values of field galaxies and clusters of
Abell richness class (ARC) 0-4 are $\sim$ 67.5, 600 $\pm$ 200, 1000
$\pm$ 200, 1400 $\pm$ 200, 1800 $\pm$ 200, and 2200 $\pm$ 200
Mpc$^{1.77}$ , respectively (the field $B_{gc}$-value is from Davis \&
Peebles 1983; the values for ARC 0-4 are from Yee \& L\'opez-Cruz
1999). The average clustering around the local ULIRGs therefore
corresponds to an environment similar to the field.  A large scatter
is seen in our data: although most objects are consistent with no
galaxy enhancement, a few objects apparently lie in clusters of Abell
classes 0 and 1.

Before discussing the results any further, it is important to verify
that our analysis of the cropped (FOV $<$ 1 Mpc) images does not
introduce any bias when compared with the results from the uncropped
(FOV $\ge$ 1 Mpc) images.  The average (median) $B_{gc}$-value for the
44 objects with uncropped images is 4 $\pm$ 121 Mpc$^{1.77}$ ($-$5
Mpc$^{1.77}$); {\em i.e.} slightly smaller than the values found for
the entire sample. Statistical tests give mixed results regarding the
significance of this discrepancy (Table 2). The results from a
two-sided K-S (Kolmogorow-Smirnov) test, a Wilcoxon
matched-pairs signed-ranks test, and a Student's t-test on the means
of the distributions suggest that the distribution of $B_{gc}$-values
for the uncropped images is not significantly different from the
distribution of $B_{gc}$-values as a whole, while the results from a
F-test on the standard deviations of the distributions suggest a
significant difference.

We have examined the distributions of $B_{gc}$-values for cropped and
uncropped images as a function of Galactic latitudes and longitudes.
Assuming that the $B_{gc}$-values are unaffected by stellar
contaminants from our Galaxy, there should be no trend with Galactic
latitude or longitude.  Indeed, the distributions of $B_{gc}$-values
with latitude and longitude are consistent with being random.  Our
data therefore confirm the results of Yee \& L\'opez-Cruz (1999), who
found that changing the counting radius by a factor of two, both
increasing to 1 Mpc and decreasing to 0.25 Mpc, did not alter the
$B_{gc}$-values significantly. However, given the mixed results from
the statistical tests, we track the cropped and uncropped data using
different symbols in the various figures of this paper. We have
verified that none of the results discussed below are affected if the
counting radius is chosen to be 0.25 Mpc or 1 Mpc rather than 0.50
Mpc, although quantization errors becomes noticeable when the counting
radius is 0.25 Mpc due to poorer number statistics. A counting radius
of 0.5 Mpc is adopted in the rest of this paper to match that of
previous studies.

Next, we explore the possibility of a dependence of the environment
richness on ULIRG properties. The first parameter we examine is the
infrared luminosity (Fig. 1). Statistical tests indicate that no trend
is present between $B_{gc}$ and $L_{IR}$. This is true for both the
uncropped data and the entire sample. The same result is found when we
examine the environment richness as a function of redshift (Fig. 2).
Here, however, the redshift range covered by our ULIRGs is very narrow
($z$ = 0.1 -- 0.22, if we exclude four objects in the sample), so this
statement is not statistically very significant. Comparisons with the
results of Blain et al.\ (2004) and Farrah et al.\ (2004, 2006)
suggest that the environment of high-$z$ ULIRGs is richer than that
of local ULIRGs. We return to this point in \S 5. 

In Figure 2, we also distinguish between optical spectral types.  We
separate our sample into Seyfert 1s, Seyfert 2s, LINERs, and HII
region-like galaxies based on the optical classification of Veilleux
et al. (1999a). No obvious trends are observed with spectral type, but
the subdivision of our sample into four subsets necessarily leads to
poorer statistics. In Figure 3, we plot $B_{gc}$ as a function of
log($f_{25}/f_{60}$), another clear indicator of AGN activity [objects
with log($f_{25}/f_{60}$) $>$ --0.7 have ``warm'', AGN-like {\em IRAS}
colors].  The lack of trends in this figure and Figure 2 indicates
that the nature of the dominant energy source in local ULIRGs
(starburst or AGN) is not influenced by the environment. This result
is consistent with the ULIRG -- QSO evolutionary scenario of Sanders
et al.\ (1988), where the nature of the dominant energy source varies
with merger phase (starburst in early phases and QSO in late phases)
but is independent of the environment (as long as the dispersion in 
velocity of the galaxies within the cluster is not too large to
prevent mergers altogether).

\section{Comparison with AGN Samples}

In this section we compare our results with those from published
environmental studies of AGNs and QSOs.  Table 2 summarizes the
statistical results of these comparisons and Figures 4 -- 9 display
the $B_{gc}$-values from the various samples. Unless otherwise noted
in the text below, all data sets use the same cosmology.

\subsection{Local Seyferts}

First, we compare our results with those derived on nearby AGNs.  De
Robertis et al.\ (1998) studied the environments of nearby ($z <
0.05$) Seyfert galaxies using the exact same procedure as the one we
use here, so we can directly compare their results with ours. For the
27 galaxies with z $>$ 0.0045, de Robertis et al.\ (1998) find
$\langle B_{gc} \rangle = 40 \pm 63$ Mpc$^{1.77}$ (median of 27
Mpc$^{1.77}$), consistent with the environment of field galaxies.
Recent studies based on the SDSS database confirm this result (e.g.,
Miller et al.\ 2003; Wake et al.\ 2004). The average environment of
local ULIRGs is therefore not dissimilar to that of local Seyferts.
However, as indicated in Table 2, virtually all statistical tests
except perhaps the t-test on the means indicate that the two $B_{gc}$
distributions are not drawn from the same parent population.  Figures
4 and 5 show why that is the case: The distribution of $B_{gc}$-values
among ULIRGs is distinctly broader than that of the Seyferts. This
slight difference is also seen in Figure 6, where we display the
distribution of local Seyferts and local ULIRGs as a function of Abell
richness classes. 

As discussed in \S 4, Seyfert-like ULIRGs do not reside in distinctly
poorer or richer environments than non-Seyfert ULIRGs, so the broader
scatter in ULIRG environments cannot be attributed to the broad range
of AGN activity level within the ULIRG population. We note that
typical error bars for the de~Robertis et al. (1998) sample is $\sim$
100 Mpc$^{1.77}$, while it is $\sim$ 150 Mpc$^{1.77}$ for the ULIRG
sample. (The difference is due to the redshift difference between the
samples -- the ULIRG data requires counting to a fainter magnitude,
which introduces larger uncertainties from background counts.) But the
full ULIRG sample distribution is about 3 times broader than the
Seyfert distribution -- so, the broader distribution of the ULIRG
sample cannot be fully explained by the larger error bars.

\subsection{Local QSOs}

Next, we compare our results with those derived on the nearby ($z
\approx 0.2$) QSOs by Yee \& Green (1984) and McLure \& Dunlop (2001).
The measurements of Yee \& Green (1984) can be directly compared with
our ULIRG results since their results were derived using the same
method and parameters as that of the present study.  McLure \& Dunlop
also apply the same formalism to calculate $B_{gc}$. However, they use a
different analysis package to identify and classify the objects in the
field and carry out the photometry. Their use of {\em HST} WFPC2 data
also limits their survey area to only $\sim$ 200 kpc around the QSOs,
smaller than even our cropped data.  These possible caveats should be
kept in mind when comparing their results with ours.

Yee \& Green (1984) get $\langle B_{gc} \rangle = 157 \pm 208$ and a
median of 134 Mpc$^{1.77}$ for 34 QSOs from the Palomar-Green sample
(Schmidt \& Green 1983), while McLure \& Dunlop (2001) derive an
average (median) $B_{gc}$ of 365 $\pm$ 404 Mpc$^{1.77}$ (241
Mpc$^{1.77}$) for a set of 44 radio-quiet and radio-loud QSOs and
radio galaxies. If we limit our discussion to the QSOs in McLure \&
Dunlop sample (21 radio-quiet QSOs and 13 radio-loud QSOs), the
average (median) $B_{gc}$ becomes 304 $\pm$ 350 Mpc$^{1.77}$ (218
Mpc$^{1.77}$). The average environment of the QSOs in both studies is
therefore slightly richer than that of local ULIRGs.  The $B_{gc}$
distributions of the two sets of local QSOs (particularly that of the
McLure \& Dunlop sample; see Fig. 5) show a distinct tail at high
$B_{gc}$-values which is not apparent in the ULIRG distribution.

A quantitative analysis generally confirms that the $B_{gc}$
distributions of Yee \& Green and $B_{gc}$ distributions for the
radio-quiet and radio-loud QSOs from McLure \& Dunlop are
statistically different from that of the local ULIRGs (Table 2).
However, note that the Wilcoxon test suggests that the difference is
barely significant.  Indeed, Figures 5, 7, and 8 show that there is
considerable overlap between the B$_{gc}$ distributions of 1-Jy ULIRGs
and low-$z$ QSOs, particularly the PG~QSOs.  This result is consistent
with the idea that some, but perhaps not all, of these QSOs may have
formed through a IR-luminous phase like that observed at low redshift
in the 1-Jy ULIRGs.  A more physically meaningful test of this
scenario would be to compare the environment of local QSOs with the
environment of $z \ga 0.5$ ULIRGs to take into account the finite
duration of the ULIRG -- QSO evolutionary sequence. The recent
environmental studies of distant ULIRGs by Blain et al.\ (2004) and
Farrah et al.\ (2004, 2006) indeed point to slightly richer
environments, which more strongly resemble the environments of the
QSOs from McLure \& Dunlop.

{\em A posteriori}, the distinct high-$B_{gc}$ tail in the
distribution of the QSOs of McLure \& Dunlop (2001) is not
unexpected given the host properties of these particular QSOs: $\sim$
4-5 times larger host sizes and luminosities relative to the 1-Jy
ULIRGs (Dunlop et al.  2003; Veilleux et al.\ 2002, 2006). More
luminous hosts live in richer environments on average than hosts of
lower luminosity. As pointed out by Veilleux et al.  (2006) and Dasyra
et al. (2006c), the hosts of the QSOs from the Palomar-Green sample
(these QSOS are less radio and X-ray luminous than the QSOs of McLure
\& Dunlop 2001) are a better match in host size and luminosity to the
local ULIRGs. This may explain the generally better (although not
perfect) agreement between the environments of PG~QSOs and 1-Jy
ULIRGs.

\subsection{Intermediate-Redshift QSOs}

For the sake of completeness, we display in Figures 5, 6, and 9 the
results from our study of local ULIRGs alongside the results presented
by Ellingson et al.\ (1991), Wold et al.\ (2001), and Barr et al.\
(2003) for 63 radio-quiet and radio-loud QSOs at 0.3 $< z < 0.6$, 20
radio-quiet QSOs at 0.5 $\le z \le 0.8$, and 20 radio-loud QSOs at 0.6
$< z <$ 1.1, respectively. All three groups use the same basic method
outlined in \S 3 to calculate the spatial correlation amplitude, and
all groups assume the same value for H$_{0}$.  However, Ellingson et
al. (1991) assume $q_0$ = 0.02 instead of 0.5 ($\Omega_m$ = 0.04
instead of 1, if $\Omega_{\lambda}$ = 0).  There is no simple way to
scale the $B_{gc}$-values for different cosmological models (other
than $H_0$) since its computation is rather complicated (\S 3), so
Figures 5, 6, and 9 show the $B_{gc}$-values corrected for the
different $H_0$ but not the different $\Omega_m$.  Once again, we see
considerable overlap between the various distributions, 
but the statistical analysis formally rules out that they come
from the same parent population (Table 2). The amount of overlap in
$B_{gc}$-values is quite remarkable given the difference in redshifts
between the various samples. These results further support a
connection between ULIRGs and some QSOs.

\section{CONCLUSIONS}

We have derived the spatial cluster-galaxy correlation amplitude,
$B_{gc}$, for 76 $z < 0.3$ ULIRGs from the 1-Jy sample and compared
our results with those in the literature on $z < 0.05$ AGNs, $z
\approx 0.2$ QSOs, and $0.3 \la z \la 1$ QSOs.  The main results are
as follows:

1. Local ULIRGs live in environments which are similar on average to
that of field galaxies. However, there are a few exceptions: some
objects apparently lie in clusters of Abell classes 0 and 1.

2. The infrared luminosity, optical spectral type, and {\em IRAS}
25-to-60 $\mu$m flux ratios of ULIRGs show no dependence with
environment.

3. The ULIRG environment does not vary systematically over the
redshift range covered by our sample (mostly 0.1 $< z <$ 0.22).

4. There is a lot of overlap between the $B_{gc}$ distribution of
local ULIRGs and those of local Seyferts, local QSOs, and
intermediate-$z$ QSOs. However, quantitative statistical comparisons
show that the various $B_{gc}$ distributions are not drawn from the
same parent population.  The average environment of ULIRGs appears to
be intermediate between that of local Seyferts and local QSOs.  Local
ULIRGs show a broader range of environments than local Seyferts, which
are exclusively found in the field. The $B_{gc}$ distribution of QSOs
show a distinct tail at high values which is not seen among local
ULIRGs. This slight environmental discrepancy between local QSOs and
ULIRGs is not unexpected: recent morphological studies have found that
some of the more radio and X-ray luminous local QSOs used in this
comparison have more luminous and massive hosts than local ULIRGs.  A
better match in host and environmental properties is seen when the
comparison is made with the PG~QSOs.

5. Overall, the results of this study suggest that ULIRGs can be a
phase in the lives of all types of AGNs and QSOs, but not all
moderate-luminosity QSOs may have gone through a ULIRG phase.
Published studies of the environments of more distant ULIRGs, perhaps
the actual predecessors of the local QSOs we see today, provide
further support for an evolutionary connection between ULIRGs and
QSOs.

\acknowledgements 

B.A.Z. and S.V. acknowledge partial support of this research by
NSF/CAREER grant AST-9874973 and NASA grant \#1263752 issued by
JPL/Caltech..  This research has made use of the NASA/IPAC
Extragalactic Database (NED) which is operated by the Jet Propulsion
Laboratory, California Institute of Technology, under contract with
the National Aeronautics and Space Administration.

\clearpage

\begin{deluxetable}{lrrccccrrr}
\tabletypesize{\tiny}
\tablecaption{Galaxies Properties}
\tablewidth{0pt}
\tablenum{1}

\tablehead{\\
\multicolumn{1}{c}{Name} & 
\multicolumn{1}{c}{$l$}  &
\multicolumn{1}{c}{$b$}  &
\multicolumn{1}{c}{$z$}  &
\multicolumn{1}{c}{log(${L_{IR}}\over{L_{\odot}}$)} &
\multicolumn{1}{c}{ST}   &
\multicolumn{1}{c}{log(${{f_{25}}\over{f_{60}}}$)} &
\multicolumn{1}{c}{B$_{gc}$}  &
\multicolumn{1}{c}{1-$\sigma$} &
\multicolumn{1}{c}{Field}
\\
\\
\multicolumn{1}{c}{(1)}  &
\multicolumn{1}{c}{(2)}  &
\multicolumn{1}{c}{(3)}  &
\multicolumn{1}{c}{(4)}  &
\multicolumn{1}{c}{(5)}  &
\multicolumn{1}{c}{(6)}  &
\multicolumn{1}{c}{(7)}  &
\multicolumn{1}{c}{(8)}  &
\multicolumn{1}{c}{(9)}  &
\multicolumn{1}{c}{(10)} 
}\label{data}

\startdata
F00091$-$0738 & 95.6 & $-$68.1 & 0.118 & 12.19 & HII & $-$1.08 & $-$80 & 97 & 1240 \\
F00188$-$0856 & 100.5 & $-$70.2 & 0.128 & 12.33 & L & $-$0.85 & $-$26 & 103 & 1330 \\
F00397$-$1312 & 113.9 & $-$75.6 & 0.261 & 12.90 & HII & $-$0.74 & 25 & 137 & 2220 \\
F00456$-$2904 & 326.4 & $-$88.2 & 0.110 & 12.12 & HII & $-$1.27 & $-$34 & 94 & 1220 \\
F00482$-$2721 & 49.4 & $-$89.8 & 0.129 & 12.00 & L & $-$0.80 & 45 & 111 & 1390 \\
\\
F01004$-$2237 & 152.1 & $-$84.6 & 0.118 & 12.24 & HII & $-$0.54 & 34 & 103 & 1290 \\
F01166$-$0844 & 143.6 & $-$70.2 & 0.118 & 12.03 & HII & $-$1.01 & 70 & 109 & 1240 \\
F01199$-$2307 & 183.3 & $-$81.8 & 0.156 & 12.26 & HII & $-$1.00 & $-$26 & 112 & 1550 \\
F01298$-$0744 & 151.1 & $-$68.1 & 0.136 & 12.27 & HII & $-$1.11 & 34 & 112 & 1390 \\
F01355$-$1814 & 174.9 & $-$75.9 & 0.192 & 12.39 & HII & $-$1.07 & $-$61 & 121 & 720 \\
\\
F01494$-$1845 & 184.3 & $-$73.6 & 0.158 & 12.23 & $-$ & $-$0.93 & $-$99 & 113 & 1620 \\
F01569$-$2939 & 225.6 & $-$74.9 & 0.141 & 12.15 & HII & $-$1.09 & $-$118 & 108 & 1430 \\
F02411$+$0353 & 168.2 & $-$48.6 & 0.144 & 12.19 & $-$ & $-$0.79 & 24 & 113 & 1450 \\
F02480$-$3745 & 243.1 & $-$63.0 & 0.165 & 12.23 & $-$ & $-$1.06 & $-$70 & 114 & 1680 \\
F03209$-$0806 & 192.0 & $-$49.3 & 0.166 & 12.19 & HII & $-$0.89 & $-$71 & 115 & 1620 \\
\\
F03250$+$1606 & 168.7 & $-$32.4 & 0.129 & 12.06 & L & $-$0.96 & $-$137 & 103 & 1330 \\
Z03521$+$0028 & 188.4 & $-$38.0 & 0.152 & 12.45 & L & $-$1.10 & $-$203 & 111 & 1520 \\
F04074$-$2801 & 225.9 & $-$46.4 & 0.153 & 12.14 & L & $-$1.28 & 121 & 130 & 1520 \\
F04103$-$2838 & 226.9 & $-$45.9 & 0.118 & 12.15 & L & $-$0.53 & 31 & 103 & 1240 \\
F04313$-$1649 & 213.6 & $-$37.8 & 0.268 & 12.55 & $-$ & $-$1.16 & $-$17 & 135 & 2260 \\
\\
F05020$-$2941 & 231.5 & $-$35.1 & 0.154 & 12.28 & L & $-$1.29 & 301 & 153 & 1530 \\
F05024$-$1941 & 220.1 & $-$32.0 & 0.192 & 12.43 & S2 & $-$0.88 & $-$25 & 121 & 1800 \\
F05156$-$3024 & 233.2 & $-$32.4 & 0.171 & 12.20 & S2 & $-$1.06 & $-$3 & 116 & 1660 \\
F08201$+$2801 & 195.3 & $+$31.3 & 0.168 & 12.23 & HII & $-$0.89 & $-$173 & 123 & 650 \\
F08474$+$1813 & 208.7 & $+$34.1 & 0.145 & 12.13 & $-$ & $-$0.83 & $-$36 & 181 & 580 \\
\\
F08591$+$5248 & 165.4 & $+$41.0 & 0.158 & 12.14 & $-$ & $-$0.80 & 65 & 143 & 620 \\
F09039$+$0503 & 225.0 & $+$32.1 & 0.125 & 12.07 & L & $-$1.09 & $-$96 & 120 & 520 \\
F09539$+$0857 & 228.5 & $+$44.8 & 0.129 & 12.03 & L & $-$0.98 & $-$100 & 121 & 530 \\
F10035$+$2740 & 202.7 & $+$53.5 & 0.165 & 12.22 & $-$ & $-$0.83 & 414 & 198 & 650 \\
F10091$+$4704 & 169.9 & $+$53.2 & 0.246 & 12.67 & L & $-$1.17 & 678 & 319 & 860 \\
\\
F10190$+$1322 & 227.2 & $+$52.4 & 0.077 & 12.00 & HII & $-$0.94 & 277 & 140 & 860 \\
F10485$-$1447 & 264.6 & $+$38.7 & 0.133 & 12.17 & L & $-$0.84 & $-$45 & 119 & 550 \\
F10594$+$3818 & 180.5 & $+$64.7 & 0.158 & 12.24 & HII & $-$0.93 & $-$11 & 125 & 620 \\
F11028$+$3130 & 196.5 & $+$66.6 & 0.199 & 12.32 & L & $-$1.05 & 2 & 131 & 740 \\
F11180$+$1623 & 235.9 & $+$66.3 & 0.166 & 12.24 & L & $-$0.80 & 119 & 156 & 650 \\
\\
F11223$-$1244 & 272.6 & $+$44.7 & 0.199 & 12.59 & S2 & $-$0.98 & 35 & 136 & 740 \\
F11387$+$4116 & 164.6 & $+$70.0 & 0.149 & 12.18 & HII & $-$0.86 & 180 & 160 & 600 \\
Z11598$-$0112 & 278.6 & $+$59.0 & 0.151 & 12.43 & S1 & $-$0.80 & $-$21 & 111 & 1510 \\
F12032$+$1707 & 254.8 & $+$75.3 & 0.217 & 12.57 & L & $-$0.74 & $-$194 & 127 & 1970 \\
F12127$-$1412 & 283.4 & $+$62.0 & 0.133 & 12.10 & L & $-$0.81 & $-$121 & 117 & 550 \\
\\
F12265$+$0219 & 290.8 & $+$62.4 & 0.159 & 12.73 & S1 & $-$0.36 & $-$6 & 149 & 1570 \\
F12359$-$0725 & 295.7 & $+$63.4 & 0.138 & 12.11 & L & $-$0.95 & 192 & 163 & 560 \\
F12447$+$3721 & 127.9 & $+$80.0 & 0.158 & 12.06 & HII & $-$1.02 & $-$103 & 122 & 620 \\
F13106$-$0922 & 311.9 & $+$52.9 & 0.174 & 12.32 & L & $-$1.32 & 102 & 131 & 1680 \\
F13218$+$0552 & 324.4 & $+$67.1 & 0.205 & 12.63 & S1 & $-$0.47 & 92 & 148 & 760 \\
\\
F13305$-$1739 & 316.8 & $+$43.8 & 0.148 & 12.21 & S2 & $-$0.47 & $-$140 & 122 & 590 \\
F13335$-$2612 & 315.3 & $+$35.3 & 0.125 & 12.06 & L & $-$1.00 & $-$58 & 101 & 1300 \\
F13342$+$3932 & 88.2 & $+$74.6 & 0.179 & 12.37 & S1 & $-$0.61 & 140 & 159 & 690 \\
F13443$+$0802 & 339.6 & $+$66.6 & 0.135 & 12.15 & S2 & $-$1.13 & $-$1 & 106 & 1380 \\
F13454$-$2956 & 317.3 & $+$31.1 & 0.129 & 12.21 & S2 & $-$1.49 & 118 & 122 & 1330 \\
\\
F13469$+$5833 & 109.1 & $+$57.2 & 0.158 & 12.15 & HII & $-$1.50 & $-$134 & 128 & 620 \\
F13509$+$0442 & 338.8 & $+$62.9 & 0.136 & 12.27 & HII & $-$0.83 & 162 & 159 & 560 \\
F14053$-$1958 & 326.4 & $+$39.1 & 0.161 & 12.12 & S2 & $-$0.86 & $-$42 & 123 & 630 \\
F14060$+$2919 & 44.0 & $+$73.0 & 0.117 & 12.03 & HII & $-$1.06 & $-$84 & 115 & 490 \\
F14121$-$0126 & 341.1 & $+$54.9 & 0.151 & 12.23 & L & $-$1.10 & $-$85 & 122 & 600 \\
\\
F14197$+$0813 & 355.5 & $+$61.2 & 0.131 & 12.00 & $-$ & $-$0.76 & $-$139 & 104 & 1350 \\
F14202$+$2615 & 35.1 & $+$69.6 & 0.159 & 12.39 & HII & $-$1.00 & 6 & 130 & 630 \\
F14252$-$1550 & 334.3 & $+$40.9 & 0.149 & 12.15 & L & $-$0.70 & 80 & 144 & 600 \\
F15043$+$5754 & 94.7 & $+$51.4 & 0.151 & 12.05 & HII & $-$1.16 & 251 & 179 & 600 \\
F15206$+$3342 & 53.5 & $+$56.9 & 0.125 & 12.18 & HII & $-$0.70 & 45 & 123 & 660 \\
\\
F15225$+$2350 & 35.9 & $+$55.3 & 0.139 & 12.10 & HII & $-$0.86 & $-$64 & 120 & 570 \\
F15327$+$2340 & 36.6 & $+$53.0 & 0.018 & 12.17 & L & $-$1.12 & -8 & 103 & 90 \\
F17044$+$6720 & 98.0 & $+$35.1 & 0.135 & 12.13 & L & $-$0.55 & $-$73 & 106 & 1440 \\
F17068$+$4027 & 64.7 & $+$36.1 & 0.179 & 12.30 & HII & $-$1.04 & 1192 & 240 & 870 \\
F17179$+$5444 & 82.5 & $+$35.0 & 0.147 & 12.20 & S2 & $-$0.83 & $-$161 & 110 & 1540 \\
\\
F21208$-$0519 & 47.3 & $-$35.9 & 0.130 & 12.01 & HII & $-$0.89 & 153 & 127 & 1340 \\
F21477$+$0502 & 62.5 & $-$35.6 & 0.171 & 12.24 & L & $-$0.85 & -76 & 116 & 1660 \\
F22491$-$1808 & 45.2 & $-$61.0 & 0.076 & 12.09 & HII & $-$1.00 & 46 & 119 & 440 \\
F22541$+$0833 & 81.2 & $-$44.6 & 0.166 & 12.23 & S2 & $-$0.82 & 73 & 126 & 1620 \\
F23060$+$0505 & 81.7 & $-$49.1 & 0.173 & 12.44 & S2 & $-$0.43 & $-$221 & 116 & 1670 \\
\\
F23129$+$2548 & 97.4 & $-$32.0 & 0.179 & 12.38 & L & $-$1.35 & 133 & 137 & 1710 \\
F23233$+$2817 & 101.1 & $-$30.6 & 0.114 & 12.00 & S2 & $-$0.65 & $-$2 & 96 & 1250 \\
F23234$+$0946 & 90.9 & $-$47.4 & 0.128 & 12.05 & L & $-$1.29 & 46 & 111 & 1330 \\
F23327$+$2913 & 103.7 & $-$30.5 & 0.107 & 12.06 & L & $-$0.98 & 78 & 108 & 1150 \\
F23389$+$0300 & 91.2 & $-$55.2 & 0.145 & 12.09 & S2 & $-$0.55 & 167 & 134 & 1460 \\
F23498$+$2423 & 106.3 & $-$36.3 & 0.212 & 12.40 & S2 & $-$0.93 & 278 & 161 & 1930 \\
\enddata
\tablenotetext{\ }
{Col (1): Name from the {\em IRAS} Faint Source Database.  The prefix
Z indicates the two objects not in the Faint Source Catalog.}
\tablenotetext{\ }
{Col. (2): Galactic longitude. }
\tablenotetext{\ }
{Col. (3): Galactic latitude. }
\tablenotetext{\ }
{Col. (4): Redshift from Kim \& Sanders (1998). }
\tablenotetext{\ }
{Col. (5): Logarithm of the infrared (8--1000 $\mu$m) luminosity in
units of solar luminosity computed using the flux in all four {\em IRAS} bands
following the prescription of Kim \& Sanders (1998).}
\tablenotetext{\ }
{Col. (6): Optical spectral type from Veilleux et al.\ (1999a).}
\tablenotetext{\ }
{Col. (7): {\em IRAS} 25-to-60 $\mu$m flux ratio. }
\tablenotetext{\ }
{Col. (8): Environment richness parameter computed using PPP program,
as described in \S 3 of this paper, in Mpc$^{1.77}$.}
\tablenotetext{\ }
{Col. (9): One-sigma uncertainty on $B_{gc}$ in Mpc$^{1.77}$.}
\tablenotetext{\ }
{Col. (10): Field size in kpc. }
\end{deluxetable}

\clearpage

\begin{deluxetable}{lrrrrrrrrrrrrr}
\tablecaption{Comparisons with AGN and QSO Environmental Studies}
\rotate
\tabletypesize{\tiny}
\tiny
\tablewidth{0pt}
\setlength{\tabcolsep}{0.06in}
\tablenum{2}
\tablehead{\\
\multicolumn{1}{c}{\bf{Sample Set}} &
\multicolumn{1}{c}{\bf{N}} &
\multicolumn{1}{c}{\bf{$\langle z \rangle$}} &
\multicolumn{1}{c}{\bf{$\langle B_{gc} \rangle$}} & 
\multicolumn{1}{c}{\bf{Error}} &
\multicolumn{1}{c}{\bf{Median}} &
\multicolumn{2}{c}{\bf{KS-test}} &
\multicolumn{2}{c}{\bf{Wilcoxon}} &
\multicolumn{2}{c}{\bf{t-test}} &
\multicolumn{2}{c}{\bf{F-test}}
\\
\\
\multicolumn{1}{c}{} &
\multicolumn{1}{c}{} &
\multicolumn{1}{c}{} &
\multicolumn{1}{c}{} &
\multicolumn{1}{c}{} &
\multicolumn{1}{c}{} &
\multicolumn{1}{c}{$P_{large}$} &
\multicolumn{1}{c}{$P_{all}$} &
\multicolumn{1}{c}{$P_{large}$} &
\multicolumn{1}{c}{$P_{all}$} &
\multicolumn{1}{c}{$P_{large}$} &
\multicolumn{1}{c}{$P_{all}$} &
\multicolumn{1}{c}{$P_{large}$} &
\multicolumn{1}{c}{$P_{all}$}
\\
\multicolumn{1}{c}{(1)}  &
\multicolumn{1}{c}{(2)}  &
\multicolumn{1}{c}{(3)}  &
\multicolumn{1}{c}{(4)}  &
\multicolumn{1}{c}{(5)}  &
\multicolumn{1}{c}{(6)}  &
\multicolumn{1}{c}{(7)}  &
\multicolumn{1}{c}{(8)}  &
\multicolumn{1}{c}{(9)}  &
\multicolumn{1}{c}{(10)} &
\multicolumn{1}{c}{(11)} &
\multicolumn{1}{c}{(12)} &
\multicolumn{1}{c}{(13)} &
\multicolumn{1}{c}{(14)} 
\\
}\label{compare}

\startdata
1 Jy ULIRGs (large FOV only)$^a$          & 44 & 0.152 &  4$\pm$121 &
$\pm$18 & -5  & 1.000 &  1.000  & 1.000 & 0.717  & 1.000  &  0.356  &
1.000 &  $<$0.001   \\
\\
1 Jy ULIRGs (all)$^b$          &   76 & 0.151 & 35$\pm$198 & $\pm$15 & -3  &
1.000  & 1.000 & 0.717  & 1.000 & 0.356   & 1.000 &  $<$0.001
 & 1.000  \\
\\
\\
de Robertis et al.\ (1998) &  27 & 0.022 & 40$\pm$64 & $\pm$13
& 27  &0.031 &0.020 & 0.067 & 0.113 &0.166 &0.901 &$<$0.001  &
 $<$0.001\\
\\
Yee \& Green (1984), PG QSOs & 34           & 0.155   & 157$\pm$208  & $\pm$28
& 134   &  0.001 & 0.001   & 0.459   & 0.001   & 0.001   & 0.004   & $<$0.001   &
0.024\\
\\
McLure \& Dunlop (2001), Entire Sample & 44 & 0.194 &  365$\pm$409 & $\pm$56 & 241
& $<$0.001 & $<$0.001        & $<$0.001  & $<$0.001    & $<$0.001 &
$<$0.001      &$<$0.001       & 0.001 \\
\\
McLure \& Dunlop (2001), Radio-Quiet \& Radio-Loud QSOs  & 34 & 0.192 & 304$\pm$355 &
$\pm$61 & 218 & $<$0.001 & $<$0.001        & 0.797      &  $<$0.001
& $<$0.001     & $<$0.001 & $<$0.001          & 0.216   \\
\\
McLure \& Dunlop (2001), Radio-Quiet QSOs & 21 & 0.174 & 326$\pm$432 & $\pm$79 & 209 & $<$0.001 &
$<$0.001           & 0.006      &  0.007       & $<$0.001     & $<$0.001 &
$<$0.001          & 0.455   \\
\\
Ellingson et al.\ (1991)  & 63 & 0.435 &  121$\pm$341 & $\pm$25 & 74   & 0.017 &
0.018          & 0.150   & 0.210      &0.032  &0.065
 & $<$0.001        &$<$0.001\\
\\
Wold et al.\ (2001), Model \#1  & 20 & 0.676 &  336$\pm$343 & $\pm$42 & 203   & $<$0.001 &
$<$0.001         & $<$0.001 & $<$0.001     &$<$0.001   & $<$0.001
    &$<$0.001        & 0.512\\
\\
Wold et al.\ (2001), Model \#2  & 20 & 0.676 &  212$\pm$332 & $\pm$43 & 146   & 0.001 &
0.001         & 0.011  & 0.019     & 0.001   &0.003
    &$<$0.001        & 0.410\\
\\
Wold et al.\ (2001), Model \#3  & 20 & 0.676 &  210$\pm$365 & $\pm$43 & 129   & 0.012 &
0.021         & 0.055  & 0.079     & 0.001   &0.005
    &$<$0.001        & 0.740\\
\\
Barr et al.\ (2003)        &   20 & 0.823 &  463$\pm$677 & $\pm$143 & 347
& $<$0.001 & $<$0.001        & 0.001 & 0.002      & $<$0.001  &$<$0.001 & $<$0.001          & 0.001 \\
\enddata 
\tablenotetext{\ }
{Col. (1): Sample set used for statistical comparison.}
\tablenotetext{\ }
{Col. (2): Number of objects in the sample. }
\tablenotetext{\ }
{Col. (3): Mean redshift of sample. }
\tablenotetext{\ }
{Col. (4): Mean $B_{gc}$-value and 1-$\sigma$ scatter around the mean
  of sample in Mpc$^{1.77}$. }
\tablenotetext{\ }
{Col. (5): Root-mean square uncertainty on the mean of the $B_{gc}$-values in Mpc$^{1.77}$. }
\tablenotetext{\ }
{Col. (6): Median $B_{gc}$-value of sample in Mpc$^{1.77}$. }
\tablenotetext{\ } {Cols. (7) and (8): Results from two-sided
  Kolmogorow-Smirnov test.  Entries in col. (7) refer to comparison
  with the set of 44 ULIRGS that have an image size greater than 1 Mpc
  $\times$ 1 Mpc, while the entries in col. (8) refer to comparison
  with the entire set of 76 ULIRG images, regardless of field size.}
\tablenotetext{\ } {Cols. (9) and (10): Results from Wilcoxon
  matched-pairs signed-ranks test.  Entries in col. (9) refer to
  comparison with the set of 44 ULIRGS that have an image size greater
  than 1 Mpc $\times$ 1 Mpc, while the entries in col. (10) refer to
  comparison with the entire set of 76 ULIRG images, regardless of
  field size.}
\tablenotetext{\ } {Cols. (11) and (12): Results from Student's t-test
  on the means of the distributions. Entries in col. (11) refer to
  comparison with the set of 44 ULIRGS that have an image size greater
  than 1 Mpc $\times$ 1 Mpc, while the entries in col. (12) refer to
  comparison with the entire set of 76 ULIRG images, regardless of
  field size.}
\tablenotetext{\ } {Cols. (13) and (14): Results from F-test on the
  standard deviations of the distributions. Entries in col. (13) refer
  to comparison with the set of 44 ULIRGS that have an image size
  greater than 1 Mpc $\times$ 1 Mpc, while the entries in col. (14)
  refer to comparison with the entire set of 76 ULIRG images,
  regardless of field size.}

\tablenotetext{a}{These entries refer to the 44 ULIRGS that
  have an image size greater than 1 Mpc $\times$ 1 Mpc.}
\tablenotetext{b}{These entries refer to the entire set of 76 ULIRG
  images, regardless of field size.}
\end{deluxetable}

\clearpage

\clearpage

\begin{figure}[htp]
\centering
\caption{The environment richness parameter versus the infrared
  luminosity for local ULIRGs. Images which were cropped smaller than
  1 Mpc $\times$ 1 Mpc are noted by open circles, while the
  non-cropped images are shown as filled circles. The horizontal
  dashed line at B$_{gc}$ = 67.5 Mpc$^{1.77}$ indicates the average
  value for typical field galaxies.  The range of environment richness
  parameters for the Abell richness classes are marked, following the
  definitions of Yee \& L\'opez-Cruz (1999).  No systematic trend is
  visible between environment richness and infrared luminosity.}
\label{bgc_luminosity}
\end{figure}

\begin{figure}[htp]
\centering
\caption{The environment richness parameter versus the redshift for
  local ULIRGs.  The symbols reflect the optical spectral types of the
  ULIRGs, as listed in Veilleux et al. (1999a): circles are Seyfert 1
  galaxies, triangles are Seyfert 2 galaxies, squares are LINERs, and
  stars are HII region-like galaxies.  Open and filled symbols stand
  for cropped and uncropped images, respectively. The meaning of the
  horizontal lines is the same as that in Fig. 1. There are no
  statistically significant trends between environment richness and
  redshift or optical spectral type. }
\label{bgc_redshift}
\end{figure}

\begin{figure}[htp]
\centering
\caption{The environment richness parameter versus the logarithm of
  the {\em IRAS} 25-to-60 $\mu$m flux ratio, log($f_{25}/f_{60}$), for
  local ULIRGs.  ULIRGs with log(f$_{25}$/f$_{60}$) $>$ $-$0.7 are
  ``warm'' AGN-like systems.  The meaning of the horizontal lines and
  symbols is the same as that in Fig. 1. No systematic trend is
  visible between environment richness and the 25-to-60 $\mu$m flux
  ratio. }
\label{f25f60}
\end{figure}

\begin{figure}[htp]
\centering
\caption{ Comparison of the environment richness parameters for the
  local ULIRGs with those of $z < 0.05$ Seyfert galaxies from de
  Robertis et al.\ (1998). The meaning of the horizontal lines and
  filled and open circles is the same as that in Fig. 1.  Pentagons
  are the data from de Robertis et al. The $B_{gc}$ distribution of
  local ULIRGs is distinctly broader than that of nearby Seyferts.}
\label{compare_seyfert}
\end{figure}

\begin{figure}[htp]
\centering
\caption{ Histograms showing the distributions of environment richness
  parameters for: ($a$) local ULIRGs from this paper (entire sample);
  ($b$) local ULIRGs from this paper (uncropped data only); ($c$) $z <
  0.05$ Seyfert galaxies from de Robertis et al. (1998); ($d$) $z
  \approx 0.2$ PG~QSOs from Yee \& Green (1984); ($e$) $z \approx 0.2$
  QSOs and radio galaxies from Dunlop \& McLure (2001); ($f$) $z
  \approx 0.2$ QSOs from Dunlop \& McLure (2001); ($g$) $z \approx
  0.2$ radio-quiet QSOs from Dunlop \& McLure (2001); ($h$) $0.3 < z <
  0.6$ radio-loud and radio-quiet QSOs from Ellingson et al. (1991);
  ($i$) $0.5 \le z \le 0.8$ radio-quiet QSOs from Wold et al.  (2001;
  model \#2 of the background galaxies); and ($j$) $0.6 < z < 1.1$
  radio-loud QSOs from Barr et al.  (2003).  The results of
  statistical comparisons between these various data sets are listed
  in Table 2.  None of these data sets appears to be drawn from the
  same parent population as the local ULIRGs, although considerable
  overlap in the values of the environmental richness parameters is
  seen between the various samples, particularly the local ULIRGs
  (this paper), local Seyferts (de Robertis et al. 1998) and PG~QSOs
  (Yee \& Green 1984). }
\end{figure}
 
\begin{figure}[htp]
\centering
\caption{ Pie-chart diagrams showing the distributions of environment
  richness parameters typical of field galaxies and clusters of Abell
  richness classes 0-4 for the eleven different samples considered in
  this paper. See caption to Fig. 5 for a description of these
  samples.  The results of statistical comparisons between these
  various data sets are listed in Table 2.  None of these data sets
  appears to be drawn from the same parent population as the local
  ULIRGs, although considerable overlap in the values of the
  environmental richness parameters is seen between the various
  samples, particularly the local ULIRGs (this paper), local Seyferts
  (de Robertis et al. 1998) and PG~QSOs (Yee \& Green 1984). }
\end{figure}
 
\begin{figure}[htp]
\centering
\caption{ Comparison of the environment richness parameter for the
  local ULIRGs with the $z \approx 0.2$ PG~QSOs of Yee \& Green
  (1984).  The meaning of the horizontal lines and open and filled
  circles is the same as that in Fig. 1.  There is considerable
  overlap in the $B_{gc}$ distributions of local ULIRGs and PG~QSOs,
  although a statistical analysis between these two sets of objects
  generally indicates that they are not drawn from the same parent
  population. }
\label{compare_mclure_01}
\end{figure}

\begin{figure}[htp]
\centering
\caption{ Comparison of the environment richness parameter for the
  local ULIRGs with the $z \approx 0.2$ radio-quiet and radio-loud
  QSOs of McLure \& Dunlop (2001; the radio galaxies are not shown).
  The meaning of the horizontal lines and open and filled circles is
  the same as that in Fig. 1.  The environment of these QSOs is
  distinctly richer on average to that of the local ULIRGs, as
  confirmed in general by a more rigorous statistical analysis. }
\label{compare_mclure_02}
\end{figure}

\begin{figure}[htp]
\centering
\caption{ Comparison of the environment richness parameter for the
  local ULIRGs with the $0.3 < z < 0.6$ radio-loud and radio-quiet
  QSOs of Ellingson et al. (1991), the $0.5 \le z \le 0.8$ radio-quiet
  QSOs of Wold et al. (2001), and the $0.6 < z < 1.1$ radio-loud QSOs
  of Barr et al.  (2003).  The meaning of the horizontal lines and
  open and filled circles is the same as that in Fig. 1.  The $B_{gc}$
  distributions of these QSOS overlap considerably with that of the
  local ULIRGs, despite the significant difference in redshifts. }
\label{compare_all}
\end{figure}

\clearpage

\setcounter{figure}{0}

\begin{figure}[htp]
\centering
\includegraphics[totalheight=0.8\textheight]{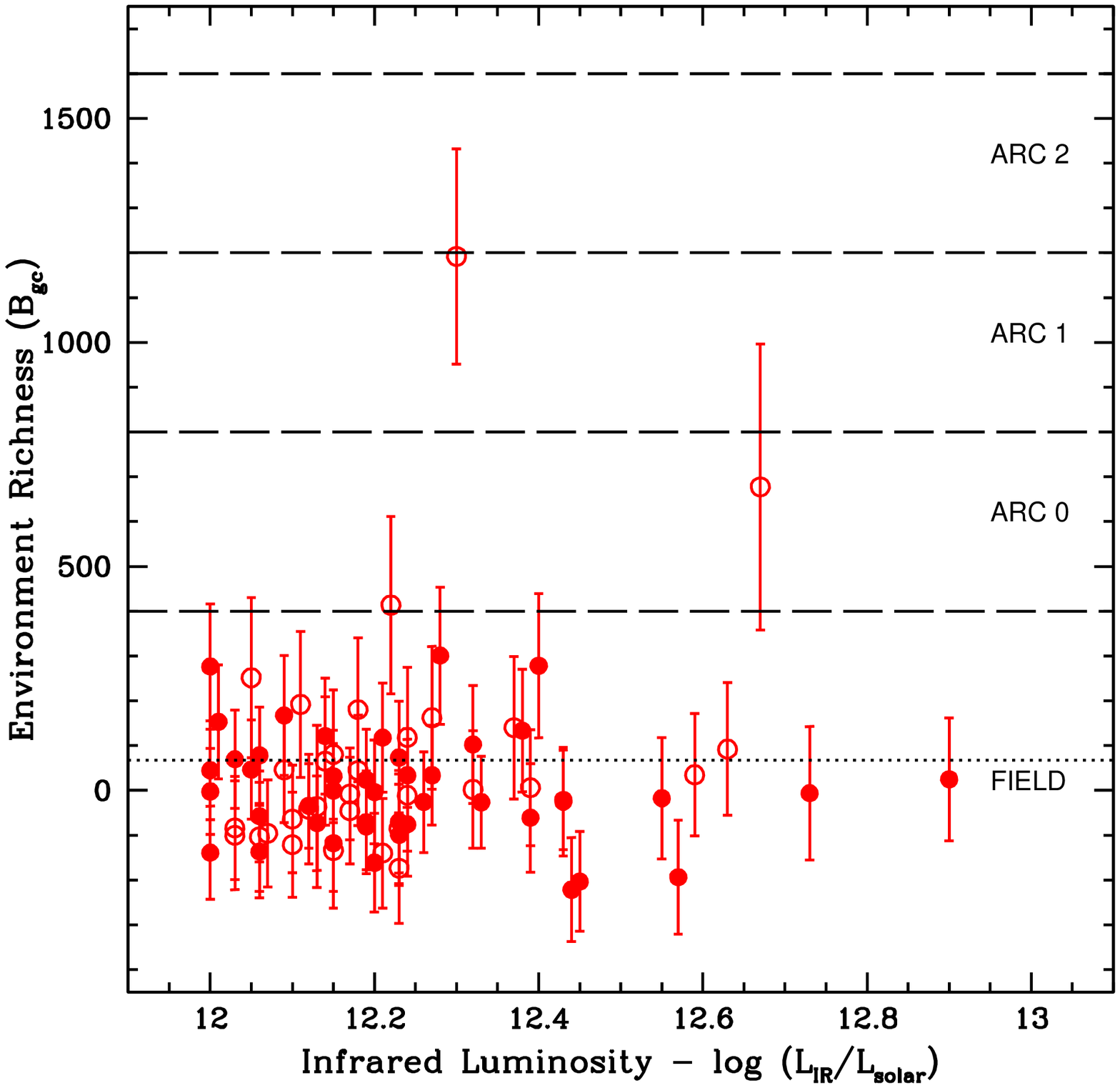}
\caption{}
\end{figure}

\begin{figure}[htp]
\centering
\includegraphics[totalheight=0.8\textheight]{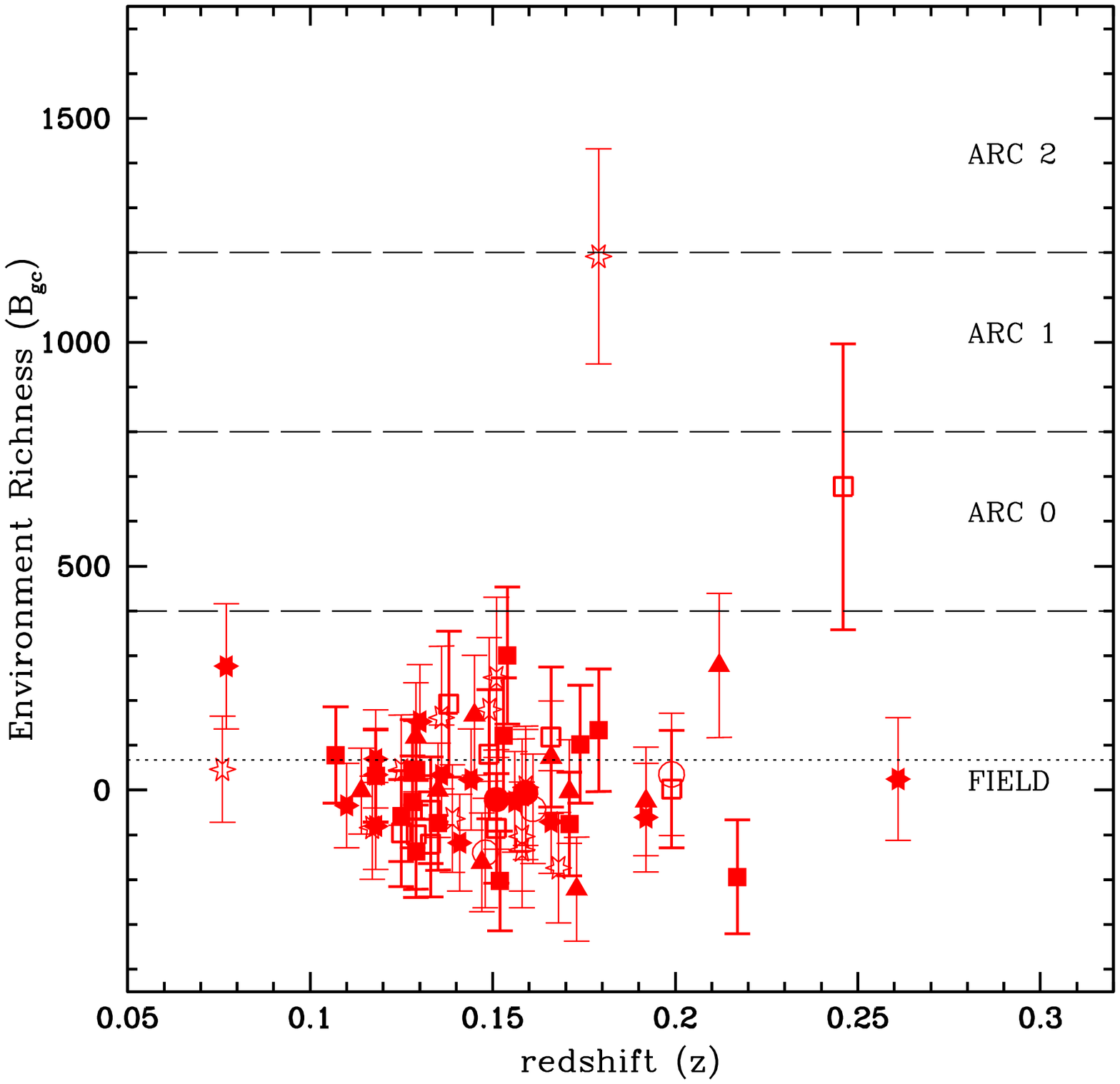}
\caption{}
\end{figure}

\begin{figure}[htp]
\centering
\includegraphics[totalheight=0.8\textheight]{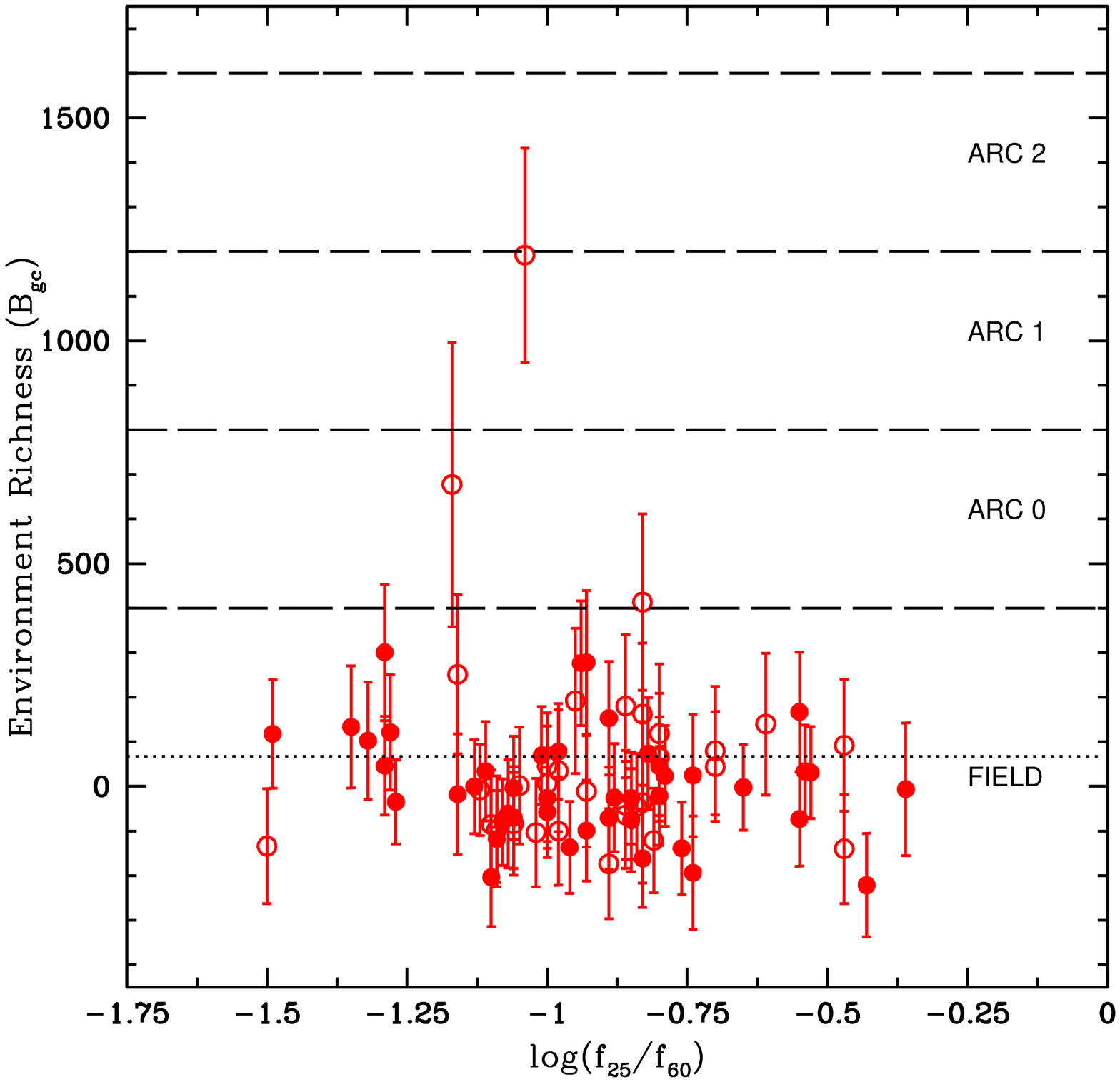}
\caption{}
\end{figure}

\begin{figure}[htp]
\centering
\includegraphics[totalheight=0.8\textheight]{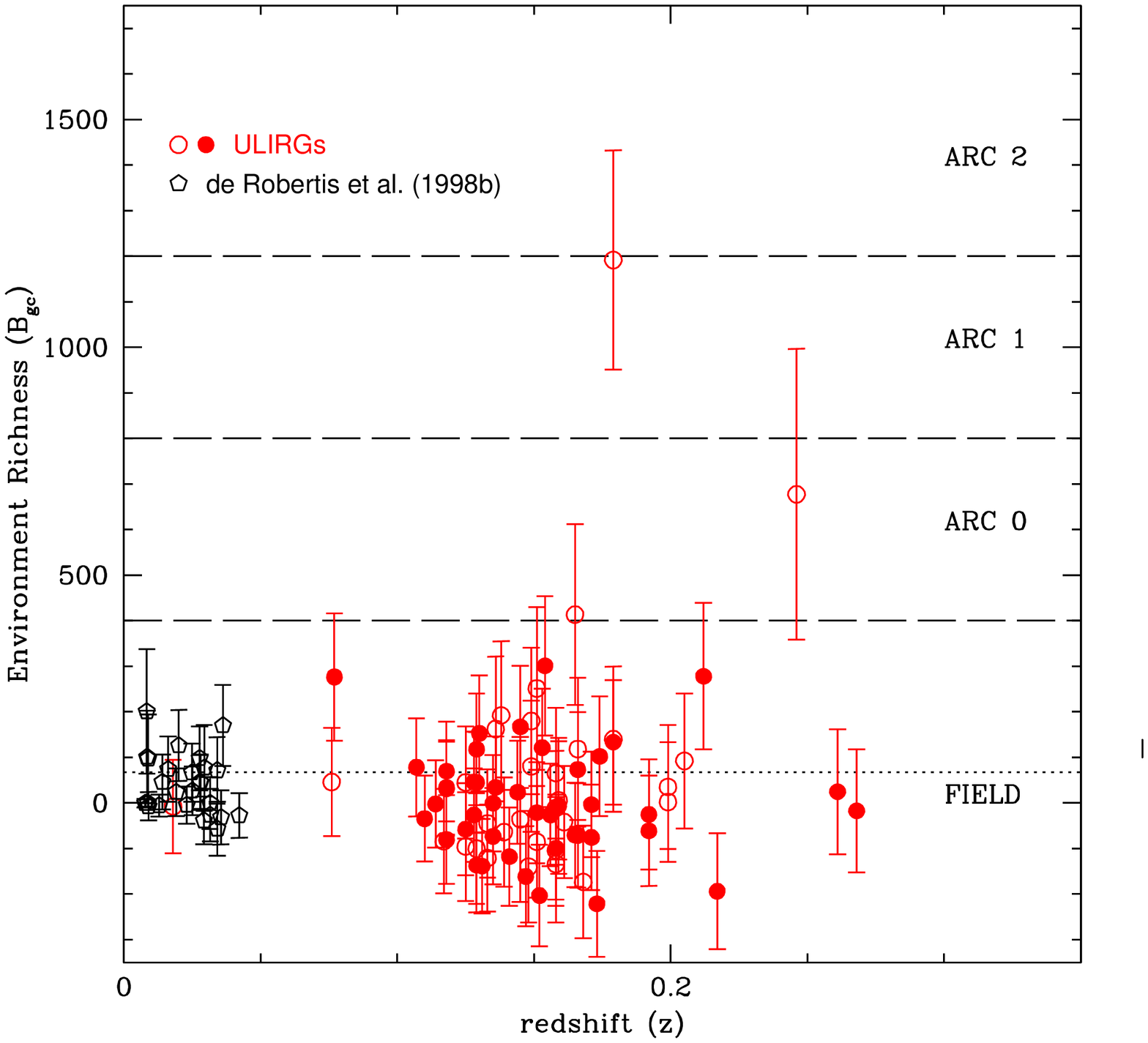}
\caption{}
\end{figure}

\begin{figure}[htp]
\centering
\includegraphics[totalheight=0.6\textheight]{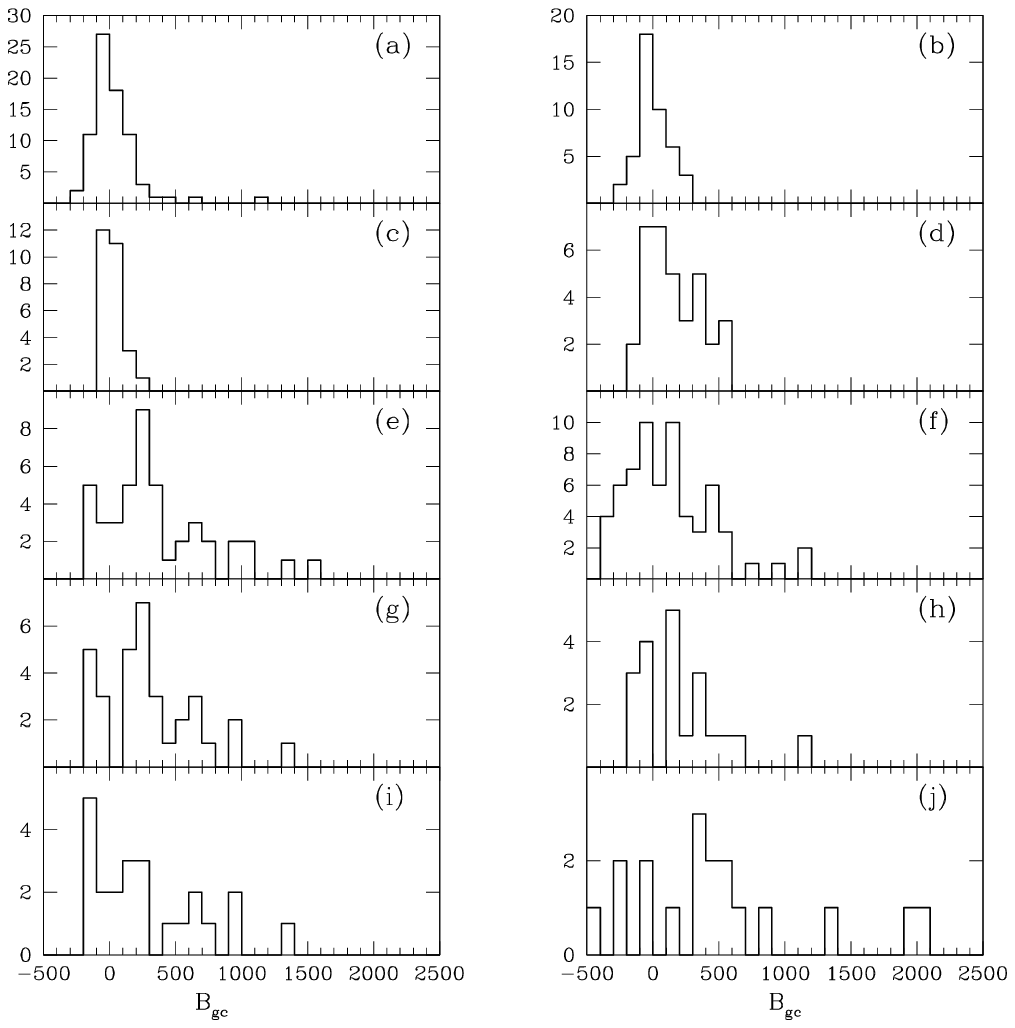}
\caption{}
\end{figure}
 
\begin{figure}[htp]
\centering
\includegraphics[totalheight=1.0\textheight]{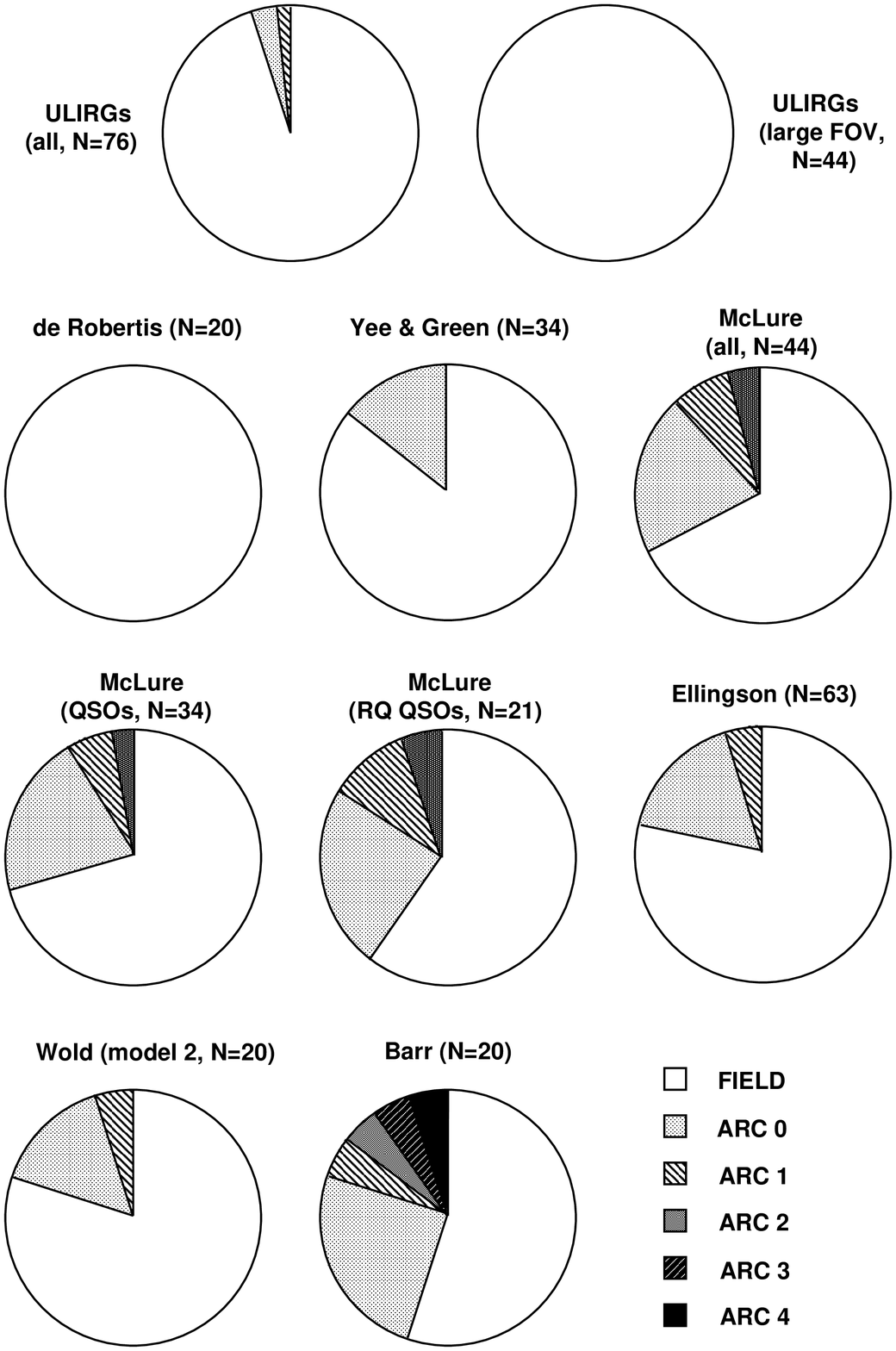}
\caption{}
\end{figure}
 
\begin{figure}[htp]
\centering
\includegraphics[totalheight=0.8\textheight]{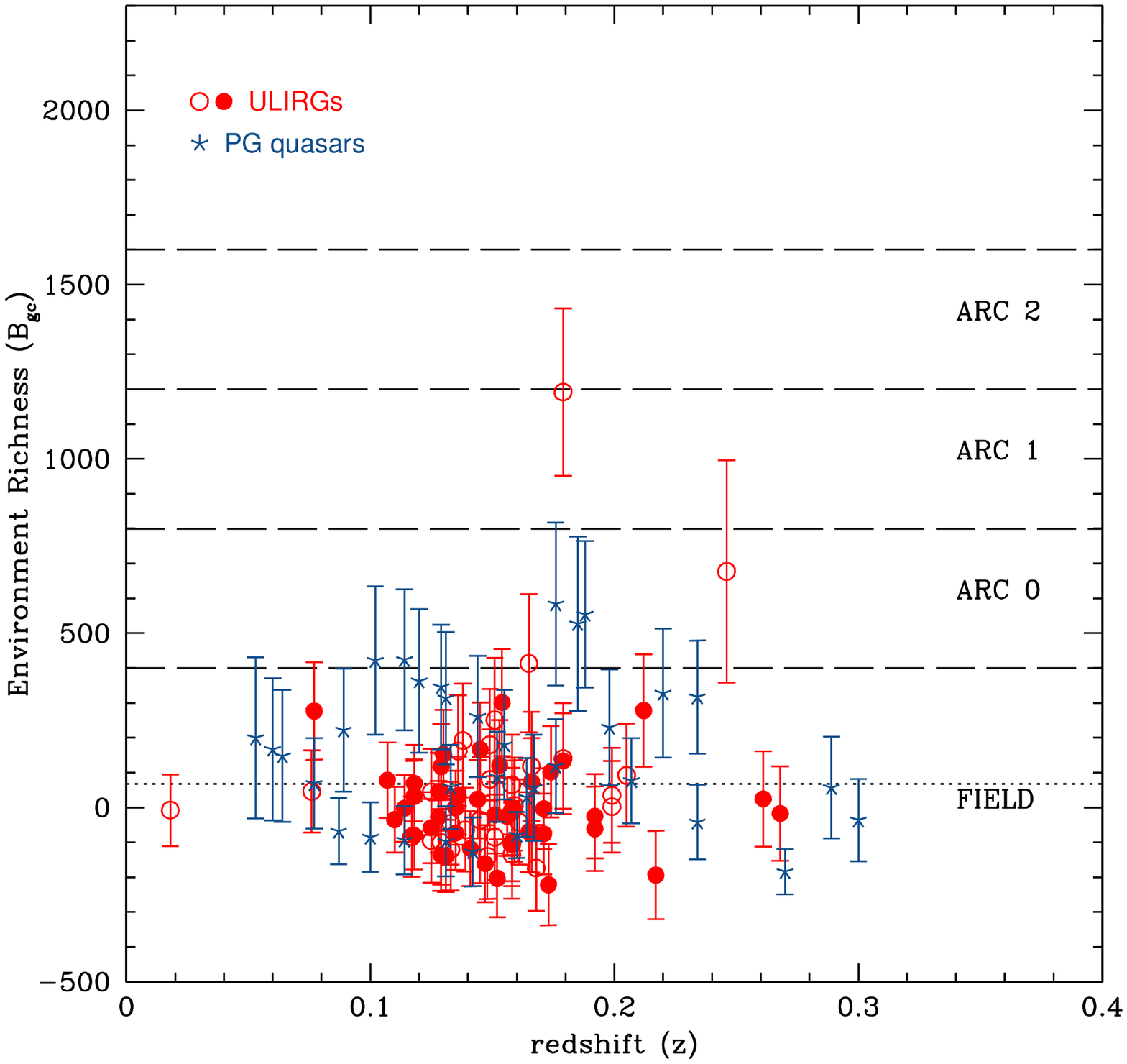}
\caption{}
\end{figure}

\begin{figure}[htp]
\centering
\includegraphics[totalheight=0.8\textheight]{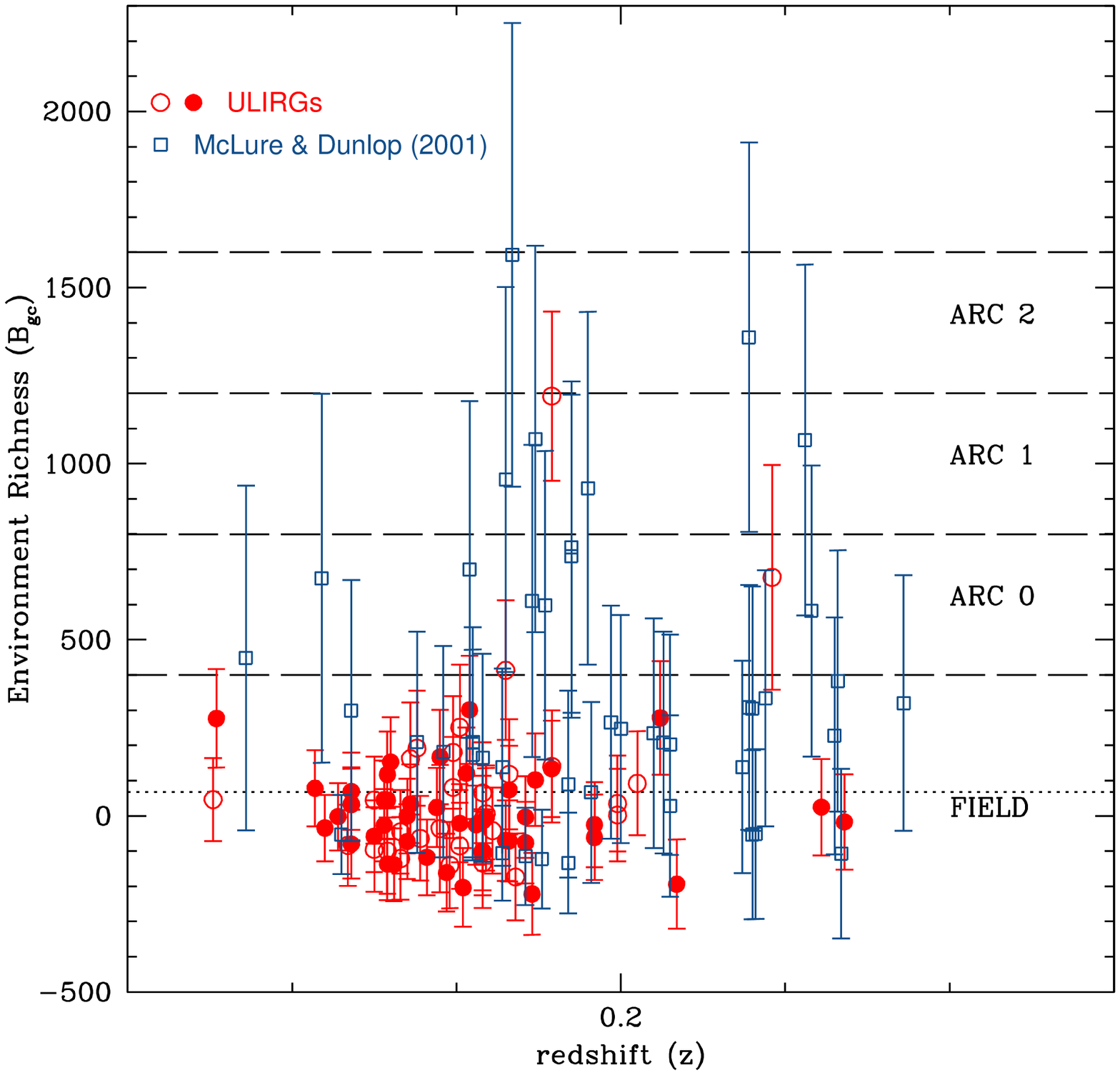}
\caption{}
\end{figure}

\begin{figure}[htp]
\centering
\includegraphics[totalheight=0.8\textheight]{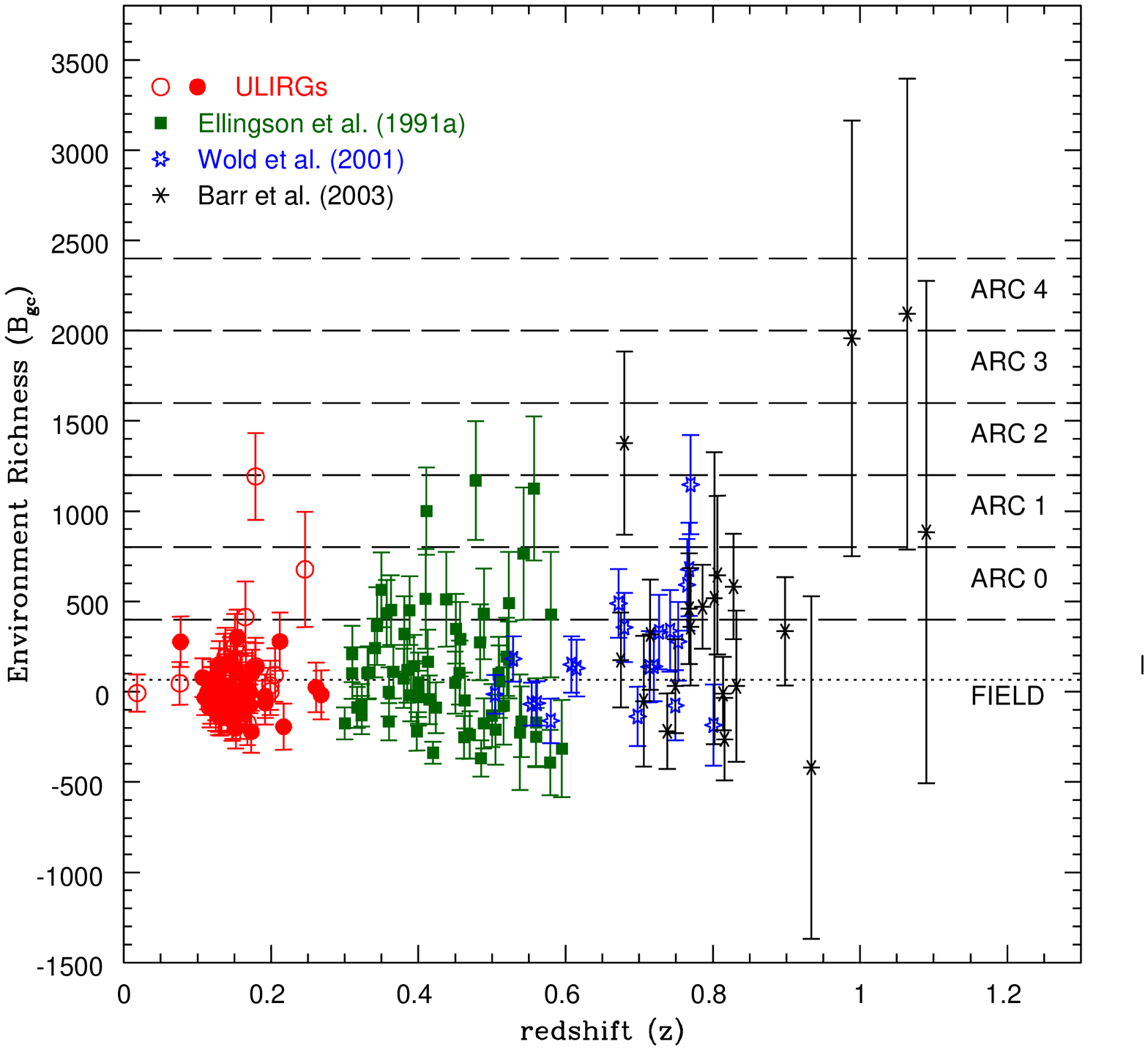}
\caption{}
\end{figure}

\clearpage


\begin{references}

\reference{}
Andersen, V., \& Owen, F. 1994, \aj, 108, 361
\reference{}
Barnes, J. E. 2004, \mnras, 350, 798
\reference{}
Barr, J. M., Bremer, M. N., Baker, J. C., \& Lehnert, M. D. 2003,
\mnras, 346, 229
\reference{}
Blain, A. W., Chapman, S. C., Smail, I., \& Ivison, R. 2004, \apj,
611, 725
\reference{}
Chapman, S. C., Blain, A. W., Smail, I., \& Ivison, R. J. 2005, \apj,
622, 772
\reference{}
Cowie, L. L., Barger, A. J., Fomalont, E. B., \& Capak, P. 2004, \apj,
603, L69
\reference{}
Croom, S. M., et al.\ 2005, \mnras, 356, 415
\reference{}
Dasyra, K., et al.\ 2006a, \apj, 638, 745
\reference{}
Dasyra, K., et al.\ 2006b, \apj, in press (astro-ph/0607468)
\reference{}
Dasyra, K., et al.\ 2006c, \apj, in press (astro-ph/0610719)
\reference{}
Davis, M., \& Peebles, P. J. E. 1983, \apj, 267, 465
\reference{}
de Robertis, M. M., Hayhoe, K., \& Yee, H. K. C. 1998a, \apjs, 115,
163
\reference{}
de Robertis, M. M., Yee, H. K. C., \& Hayhoe, K. 1998b, \apj, 496, 93
\reference{}
Downes, D., \& Solomon, P. M. 1998, \apj, 507, 615
\reference{}
Dunlop, J. S., McLure, R. J., Kukula, M. J., Baum, S. A., O'Dea,
C. P., \& Hughes, D. H. 2003, \mnras, 340, 1095 
\reference{}
Ellingson, E., Yee, H. K. C., \& Green, R. F. 1991, \apj, 371, 49
\reference{}
Farrah, D., et al.\ 2004, \mnras, 349, 518
\reference{}
Farrah, D., et al.\ 2006, \apj, 641, L17
\reference{}
Genzel, R., Tacconi, L. J., Rigopoulou, D., Lutz, D., \& Tecza,
M. 2001, \apj, 563, 527
\reference{}
Genzel, R., et al. 1998, \apj, 498, 579
\reference{}
Gladders, M. D., \& Yee, H. K. C. 2005, \apjs, 157, 1
\reference{}
Hill, G. J., \& Lilly, S. J. 1991, \apj, 367, 1
\reference{}
Hughes, D. H., et al. 1998, Nature, 394, 241
\reference{}
Kauffmann, G., et al.\ 2004, \mnras, 353, 713
\reference{}
Kim, D.-C., \& Sanders, D. B. 1998, \apjs, 119, 41
\reference{}
Kim, D.-C., Veilleux, S., \& Sanders, D. B. 1998, \apj, 508, 627
\reference{}
Kim, D.-C., Veilleux, S., \& Sanders, D. B. 2002, \apjs, 143, 277
\reference{}
Kron, R. G. 1980, \apjs, 43, 305
\reference{}
Longair, M. S., \& Seldner, M. 1979, \mnras, 189, 433
\reference{}
Lonsdale, C. J., Farrah, D., \& Smith, H. 2006, preprint (astro-ph/0603031)
\reference{}
Lutz, D., Veilleux, S., \& Genzel, R. 1999, \apj, 517, L13
\reference{}
Lutz, D., et al. 1998, \apj, 505, L103
\reference{}
Martin, C. L. 2005, \apj, 621, 227
\reference{}
McLure, R. J., \& Dunlop, J. S. 2001, \mnras, 321, 515
\reference{}
Mihos, J. C, \& Hernquist, L. 1996, \apj, 464, 641
\reference{}
Miller, C. J., Nichol, R. C., G\'omez, P. L., Hopkins, A. M., \&
Bernardi, M. 2003, \apj, 597, 142
\reference{}
Rigopoulou, D., et al. 1999, \aj, 118, 262
\reference{}
Rupke, D. S., Veilleux, S., \& Sanders, D. B. 2002, \apj, 570, 588
\reference{}
Rupke, D. S., Veilleux, S., \& Sanders, D. B. 2005a, \apjs, 160, 115
\reference{}
Rupke, D. S., Veilleux, S., \& Sanders, D. B. 2005b, \apj, 632, 751
\reference{}
Sanders, D. B., \& Mirabel, I. F. 1996, \araa, 34, 749
\reference{}
Sanders, D. B., et al.\ 1988, \apj, 325, 74
\reference{}
Schmidt, M., \& Green, R. F. 1983, \apj, 269, 352
\reference{}
Scoville, N. Z., et al. 2000, \aj, 119, 991
\reference{}
Serber, W., Bahcall, N., M\'enard, B., \& Richards, G. 2006, \apj,
643, 68
\reference{}
Smail, I., Ivison, R. J., \& Blain, A. W. 1997, \apj, 490, L5
\reference{}
S\"ochting, I. K., Clowes, R. B., \& Campusano, L. E. 2004, \mnras,
347, 1241
\reference{}
Soifer, B. T., et al. 2000, \aj, 119, 509
\reference{}
Soifer, B. T., et al. 2001, \aj, 122, 1213
\reference{}
Surace, J. A., \& Sanders, D. B. 1999, \apj, 512, 162
\reference{}
Surace, J. A., Sanders, D. B., \& Evans, A. S. et al. 2001, \aj, 122, 2791
\reference{}
Tacconi, L. J., et al.\ 2002, \apj, 580, 73
\reference{}
Tran, Q. D., et la. 2001, \apj, 552, 527
\reference{}
Veilleux, S., Cecil, G., \& Bland-Hawthorn, J. 2005, \araa, 43, 769
\reference{}
Veilleux, S., Kim, D.-C., \& Sanders, D. B. 1999a, \apj, 522, 113
\reference{}
Veilleux, S., Kim, D.-C., \& Sanders, D. B. 2002, \apjs, 143, 315
\reference{}
Veilleux, S., Sanders, D. B., \& Kim, D.-C. 1999b, \apj, 522, 139
\reference{}
Veilleux, S., et al.\ 2006, \apj, 643, 707
\reference{}
Waskett, T. J., Eales, S. A., Gear, W. K., McCracken, H. J., Lilly,
S., \& Brodwin, M. 2005, \mnras, 363, 801
\reference{}
Wold, M., Lacy, M., Lilje, P. B., \& Serjeant, S. 2000, \mnras, 316, 267
\reference{}
Wold, M., Lacy, M., Lilje, P. B., \& Serjeant, S. 2001, \mnras, 323, 231
\reference{}
Yee, H. K. C. 1991, \pasp, 103, 396
\reference{}
Yee, H. K. C., \& Ellingson, E. 2003, \apj, 585, 215
\reference{}
Yee, H. K. C., \& Green, R. F. 1984, \apj, 280, 79
\reference{}
Yee, H. K. C., \& Green, R. F. 1987, \apj, 319, 28
\reference{}
Yee, H. K. C., Green, R. F., \& Stockman, H. S. 1986, \apjs, 63, 681
\reference{}
Yee, H. K. C., \& L\'opez-Cruz, O. 1999, \aj, 117, 1985
\end{references}
\end{document}